\documentclass[lettersize,journal]{IEEEtran}
\usepackage{amsmath,amsfonts}
\usepackage{algorithm}
\usepackage{array}
\usepackage[caption=false,font=normalsize,labelfont=sf,textfont=sf]{subfig}
\usepackage{textcomp}
\usepackage{stfloats}
\usepackage{url}
\usepackage{verbatim}
\usepackage{graphicx}
\usepackage{cite}
\usepackage{float}
\usepackage{multicol}
\usepackage[table]{xcolor}
\usepackage{caption}
\usepackage{multirow}
\usepackage{adjustbox}
\usepackage{blindtext}
\usepackage{graphicx}
\usepackage{balance}
\usepackage{threeparttable}
\usepackage{algorithm}
\usepackage[noend]{algpseudocode}

\usepackage{soul}

\begin{document}

\title{Design Space Exploration of Sparsity-Aware Application-Specific Spiking Neural Network Accelerators}


\author{Ilkin Aliyev, Kama Svoboda, and Tosiron Adegbija,~\IEEEmembership{Senior Member,~IEEE} 
\thanks{The authors are with the Department of Electrical and Computer Engineering, The University of Arizona, USA, email: \{ilkina, ksvoboda, tosiron\}@arizona.edu.}
}



\maketitle
\begin{abstract}
Spiking Neural Networks (SNNs) offer a promising alternative to Artificial Neural Networks (ANNs) for deep learning applications, particularly in resource-constrained systems. This is largely due to their inherent sparsity, influenced by factors such as the input dataset, the length of the spike train, and the network topology. While a few prior works have demonstrated the advantages of incorporating sparsity into the hardware design, especially in terms of reducing energy consumption, the impact on hardware resources has not yet been explored. This is where design space exploration (DSE) becomes crucial, as it allows for the optimization of hardware performance by tailoring both the hardware and model parameters to suit specific application needs. However, DSE can be extremely challenging given the potentially large design space and the interplay of hardware architecture design choices and application-specific model parameters. 

In this paper, we propose a flexible hardware design that leverages the sparsity of SNNs to identify highly efficient, application-specific accelerator designs. We develop a high-level, cycle-accurate simulation framework for this hardware and demonstrate the framework's benefits in enabling detailed and fine-grained exploration of SNN design choices, such as the layer-wise logical-to-hardware ratio (LHR). Our experimental results show that our design can (i) achieve up to $76\%$ reduction in hardware resources and (ii) deliver a speed increase of up to $31.25\times$, while requiring $27\%$ fewer hardware resources compared to sparsity-oblivious designs. We further showcase the robustness of our framework by varying spike train lengths with different neuron population sizes to find the optimal trade-off points between accuracy and hardware latency.

\end{abstract}


\section{Introduction}
Artificial Neural Networks (ANNs) have grown exponentially in popularity, as machine learning (ML)-based methods become applicable to an increasing number of new application domains. Ever-increasing workload demands in edge computing require ANN accelerators to further reduce inference latency and energy consumption. However, ANNs are not always viable in resource-constrained systems, as they are extremely compute-intensive and can be prohibitive for edge computing applications despite their high prediction accuracy. 

Spiking Neural Networks (SNNs) are gaining a lot of attention as an efficient alternative to ANNs for machine learning in resource-constrained systems \cite{rathi2022exploring}. SNNs are a special kind of neural networks that differ from ANNs in their communication and computation schemes. Neurons in SNNs transmit discrete binary events (or \textit{spikes}) to communicate with each other, rather than continuous variables as in ANNs. Whereas neurons in ANNs require complex multiply-and-accumulation (MAC) operations, SNNs only require simple addition operations. 

Furthermore, SNNs reflect biological neural networks by implementing sparse coding \cite{foldiak2003sparse} and sparse connectivity \cite{faghihi2022neuroscience}. In sparse coding, only a fraction of neurons are activated at a time, and each neuron connects with only a subset of other neurons in sparse connectivity. This sparsity can further reduce computational complexity and decrease energy consumption compared to ANNs, particularly in handling high-dimensional data. This sparsity can be leveraged to facilitate the design of efficient hardware accelerators for ML, which are ideal for power-constrained devices such as the Internet of Things (IoT) systems and edge computing devices. Therefore, SNNs not only provide a better analog of the biological neuronal communication and computation mechanisms but also offer an excellent opportunity for hardware implementation of highly efficient machine learning accelerators.

\noindent\textbf{Major issue with current SNN accelerators}: The design of SNN architectures has been an active area of research due to the benefits of SNNs for low-overhead ML. Both industry and academia have proposed various SNN accelerators, such as IBM's Truenorth \cite{truenorth}, Intel's Loihi \cite{davies2018loihi}, Spinnaker \cite{furber2014spinnnaker}, Minitaur \cite{neil2014minitaur},  S2N2 \cite{khodamoradi2021s2n2}, etc. However, prior studies (e.g., \cite{li2022efficiency}, \cite{yin2022sata}) have demonstrated the complexity of training SNNs and shown that while SNNs can be trained to achieve similar accuracy as ANNs, this is usually at the expense of energy efficiency due to the processing time steps intrinsic to SNNs. In order to close the energy efficiency gap, SNN hardware must be carefully designed to match the application behavior and exploit such characteristics as the intensity of firing activity.
Therefore, early design space exploration methodologies are needed to investigate the application-driven hardware performance and to provide opportunities for model updates before hardware synthesis and deployment on edge devices.

\noindent\textbf{Limited work on SNN design space exploration}: Prior work on SNN design space exploration (DSE) studied the hardware efficiency implications of model parameters, but these DSE methods are limited to a small number of parameters such as spike encoding mechanisms, degree of parallelism \cite{abderrahmane2020design}, and spike train length  \cite{fang2020encoding, 2022_hw_footprint}. In contrast, we adopt an expanded view of the neuronal dynamics of SNNs and how they affect and are affected by hardware designs, especially considering the network's sparsity. We propose a cycle-accurate simulation approach for exploring various neural parameters, including the degree of parallelism, and the ratio of logical neurons to physical hardware neurons. Importantly, our approach can study the impacts of these model parameters at a fine, layer-wise granularity.

\noindent\textbf{Sparsity-aware SNN hardware}: A neuron's workload in an SNN is primarily determined by the spiking intensity (particularly the pre-synaptic layer's spikes), which is influenced by factors like the dataset/application and input encoding mechanism. A higher spiking activity results in a larger accumulation delay for post-synaptic neurons, as more neurons are activated in the pre-synaptic layer. We argue that hardware resources (i.e., neuron processor, memory blocks) can be allocated based on a layer's sparsity level, alleviating high resource demands and enabling optimal performance of the SNN models. For example, recent studies \cite{SIS_2023_CAS, SIS2022_ISCAS} have shown reductions in both hardware resources and inference time by simply considering sparsity in input layer in a two layer network (e.g., only input and output layers).

We present a highly flexible sparsity-aware and cycle-accurate simulation framework for rapidly exploring the design space of application-specific SNN accelerators. The framework leverages the \textit{Transaction-Level Modeling (TLM)} \cite{cai2003transaction} formalism, which can model complex digital systems that involve complex data communication. TLM abstracts away the communication details from those of the functional units and communication architecture. This enables an abstraction that enhances modularity, composability, reusability, and interoperability of design. We implement our framework in SystemC and validate it extensively against both software and hardware implementations of SNNs. 

Through detailed experiments and analysis using our framework, we draw two key insights that may elude state-of-the-art SNN DSE methods. First, the implications of an SNN's neural dynamics on the hardware implementations vary for different layers within a network. This insight requires exploring parameters such as the total number of memory blocks and the number of physical neuron processors per layer to improve overall network efficiency. Second, increasing the logical-to-hardware neuron ratio for the deeper layers in a deep network can reduce the hardware footprint substantially without degrading the inference latency. This insight enables deploying larger and more accurate models on hardware-limited systems. To our knowledge, we are the first to perform rapid experimentation through various model configurations for an application dataset to find the sweet spot across hardware area, latency, and model accuracy. Moreover, this is also the first time that the layer-wise dynamics and sparsity of the SNN are taken into account in the design of SNN accelerators.

In summary, this work makes the following important contributions:
\begin{itemize}
    \item We propose a modular hardware design that enables the flexibility to easily adjust the allocation of hardware neurons according to layer-specific sparsity. The proposed hardware architecture takes advantage of SNN's binary communication scheme and implements it using simple hardware primitives, like shift register, priority encoder, and concatenation.
    \item We implement a cycle-accurate simulation framework for this hardware with a high degree of automation and introduce a logical-to-hardware neuron ratio (LHR) knob which controls the total number of hardware neurons allocated to each network layer.
    \item Using three different datasets, MNIST, FashionMNIST, and DVSGesture we analyze the sparsity and show area-efficient hardware with a trade-off in inference delay.
    \item Our experiments show that compared to prior works with fixed hardware configurations, our design can achieve (i) up to $76\%$ reduction in hardware resources with similar latency for MNIST, (ii) up to $31.25\times$ speed up, while requiring $27\%$ fewer hardware resources for FashionMNIST, and (iii) $2.34\times$ speed up for DVSGesture by simply tuning the layer-wise LHR knob.  
    \item Furthermore, we employ a population of neurons for the classification layer and conduct a trade-off analysis between spike train length and population size and their impact on classification accuracy and hardware performance.
    \item Finally, we open-source our code to flourish research in the area https://github.com/githubofaliyev/SNN-DSE
\end{itemize}


\section{Background and Related Work}
In this section, we briefly describe SNNs and discuss some related work on SNN design space. We then motivate and describe TLM, which we leverage in our work. We also explain its abstraction levels and their associated characteristics and review prior work on TLM-based architecture modeling. 

\subsection{Overview of Spiking Neural Networks}
SNNs are inspired by how neurons in the brain communicate via sparse, discrete electrical signals, or spikes\cite{kandel2000principles}. Modeled after the structure and functionality of biological neurons, the neurons in a typical SNN operate as simple integrate-and-fire units. This means that they accumulate incoming spikes over time and emit an outgoing spike when the integrated value reaches a certain threshold \cite{maass1997networks}. A sequence of spikes forms a spike train. Information is relayed through these spike trains via various coding schemes: rate coding, which concerns the frequency of transmitted spikes transmitted \cite{adrian1926impulses}, temporal coding or TTFS coding, which focuses on the timing of the spikes, often in relation to the time-to-first-spike \cite{johansson2004first}, burst coding, which counts the number of spikes and the inter-spike interval within a burst of spikes \cite{izhikevich2003bursts}, or phase coding, which encodes information in the spike times relative to an oscillatory background activity \cite{kim2018deep}.

A major advantage of using SNNs over traditional ANNs is their event-driven communication: they only communicate when necessary rather than continuously. This means that the neurons in SNNs only activate when a spike is present \cite{arbib2003handbook}. Moreover, unlike ANNs that require hardware-expensive Multiplier and Accumulate (MAC) operations for regular computations, SNNs can rely solely on accumulate operations \cite{rathi2021diet}. Therefore, SNNs have lower power requirements and computational costs than ANNs, which makes them ideal for edge computing applications on resource-constrained devices \cite{rueckauer2017conversion}.


\subsection{Prior work on SNN Hardware Design Space Exploration}
Despite the computational simplicity of spiking neurons, a few recent studies have argued that SNNs require higher energy and longer inference latency to achieve \textit{similar classification accuracy} to ANNs \cite{li2022efficiency, 2022_hw_footprint, yin2022sata}. While these studies may highlight the challenge of designing SNNs with comparable accuracy to ANNs, it is critical to note that these findings underscore the need for more in-depth and targeted exploration of SNNs' design space with a focus on hardware-software co-design. Given the complexity and breadth of SNN accelerators' design space---encompassing factors like network topologies, memory configurations, parallelization of computation resources, neuronal dynamics---it is clear that innovative approaches are essential for efficient DSE in SNNs, to foster the creation of effective and highly-tailored accelerators.

Li et al. \cite{li2022efficiency} compared convolutional neural network (CNN) accelerators with their spike-coding equivalents (SNNs) in terms of processing and energy efficiency using high-level synthesis (HLS) to generate CNNs and SNNs on field-programmable gate arrays (FPGAs). The study used three types of deep neural network accelerators: CNN hardware generated with HLS, SNN hardware generated with HLS, and SNN RTL hardware manually developed in VHDL. They evaluated all three accelerator configurations with the same layer-based architecture across three benchmark datasets: MNIST, GTSRB, and CIFAR-10. 

The authors found that SNNs offer comparable accuracy but may be less efficient than CNNs in terms of execution time due to the spike encoding scheme and the lack of parallelism. They used rate coding as the spike encoding scheme, which results in relatively larger spike trains and higher activity, leading to long execution times. However, the authors do not measure how many time steps are needed for SNNs to match ANN accuracy. \cite{2022_hw_footprint} study spiking activity per layer to find (i) spike train length and (ii) hardware resources needed. For example, they suggest fully parallel and flat serial hardware. They show that parallel SNNs use less energy than parallel ANNs for a complex dataset like Spoken MNIST. However, serial SNNs use more energy than serial ANNs

Another work by the authors \cite{abderrahmane2020design} examines two main parameters in SNN design: input data encoding and parallelism degree. They propose three configurations: fully parallel, time-multiplexed, and hybrid. In the fully parallel one, each logical neuron has a physical neuron. In the time-multiplexed one, one hardware neuron serves a whole layer. In the hybrid one, the first hidden layer is fully parallel, but the rest are time-multiplexed. However, both works only consider flat parallelization or serialization of layers without exploring sparsity per layer. Unlike these prior works, our framework enables better flexibility and allows users to explore the layer-wise resource allocation scheme at a finer granularity, providing more control over the trade-offs in the exploration process and the output design efficiency. Moreover, our approach streamlines the design process by letting users specify parameters such as the total number of neurons and memory blocks per layer. The framework then automatically performs the mapping of the corresponding hardware neurons. Therefore, we can change the architectural configuration easily, allowing rapid pruning of optimal hardware that matches the neural structure for the target application.

The authors in \cite{SIS2022_ISCAS} and \cite{SIS_2023_CAS} investigate how sparsity affects FPGA hardware resources and inference time. Both works use a Selective input Sparsity approach \cite{SIS2022_ISCAS} on a two-layer MLP network and present quantitative analyses. Across different datasets, results show that, at the cost of lower accuracy, a sparse connection reduces hardware area and inference time compared to a full connection. In our work, we do not use any selection mechanism but simple hardware logic to compress spike trains and remove non-spiking outputs from pre-synaptic neurons. As such, our approach does not change network accuracy. In addition, unlike prior work, our approach also applies to hidden layers.

\subsection{Overview of Transaction-Level Modeling}
Transaction-level models (TLMs) \cite{cai2003transaction} model the hardware system components at a high level of abstraction in which the details of communication among computation units are separated from the details of the computation units. Channels model communication. Transaction requests call interface functions of these channel models. The fundamental purpose of TLM is to abstract away the unnecessary details of communication and computation to speed up the simulation and enable the exploration and validation of design alternatives at a higher level of abstraction. The TLM formalism is especially suitable for simulating SNNs because of the complexity of the event-driven communications between their components. The separation and abstraction enhance modularity, composability, reusability, and interoperability of design. That is, atomic computation and communication components of an SNN (e.g., neuron, synaptic connections) can be individually simulated and validated. The component designs can be coupled to form complex systems. These designs can also be reused or coupled with designs from different vendors or designers to create new application- or domain-specific system designs. TLM also supports simulation at different levels of abstraction to allow hardware designers to explore hardware at a range of granularities. Supported abstraction levels range from the \textit{specification model}, which focuses on event ordering similar to dataflow computation without delving into computation or communication component specifics, to the \textit{component-assembly} and \textit{bus-functional} models, which introduce details of processing elements and connecting buses, respectively. Lastly, the \textit{implementation model}, which we use in our work, is the least abstract and delivers cycle-accurate modeling for both computations and communications, detailing computation tasks at the RTL granularity.

\subsection{TLM Architecture Modeling}

Embedded systems are one of the major application areas of TLMs because they contain multiple processor cores, memory/cache subsystems, and various I/O peripheral units. By enabling the rapid simulation of different models, TLM provides a quick and iterative design scheme during the early design stage of the embedded system development. The simulation time required for TLM models varies from around 1/1000th to 1/100th of the execution time of RTL design \cite{pasricha2002transaction}. TLM has the significant advantage of having multiple abstraction levels. Once the architectural specification is defined, software developers can start building their TLM models without waiting for RTL development kick-off. Consequently, TLM models can save orders of magnitude in man-hours and development costs compared to the traditional development cycle. For example, STMicroelectronics' System Architecture group (CR\&D) used TLM models for developing MPEG4 IVT six months before the top-level netlist was made available \cite{clouard2002towards}. Besides providing fast simulation, the fidelity of the TLM models has also been investigated. The study in \cite{clouard2002towards} compares TLM and RTL implementations of a dual-core processor and found that the TLM model had less than a 15\% error margin for interrupt latency and bus utilization. Although we are the first to employ TLM in modeling and simulating SNN accelerators, we envision that this approach will become a mainstay in designing and developing application-specific SNN accelerators in both industry and academia because of its numerous benefits to the design process.

\begin{figure}[t!]
\vspace{-15pt}
		\centering
		\includegraphics[width=1\linewidth]{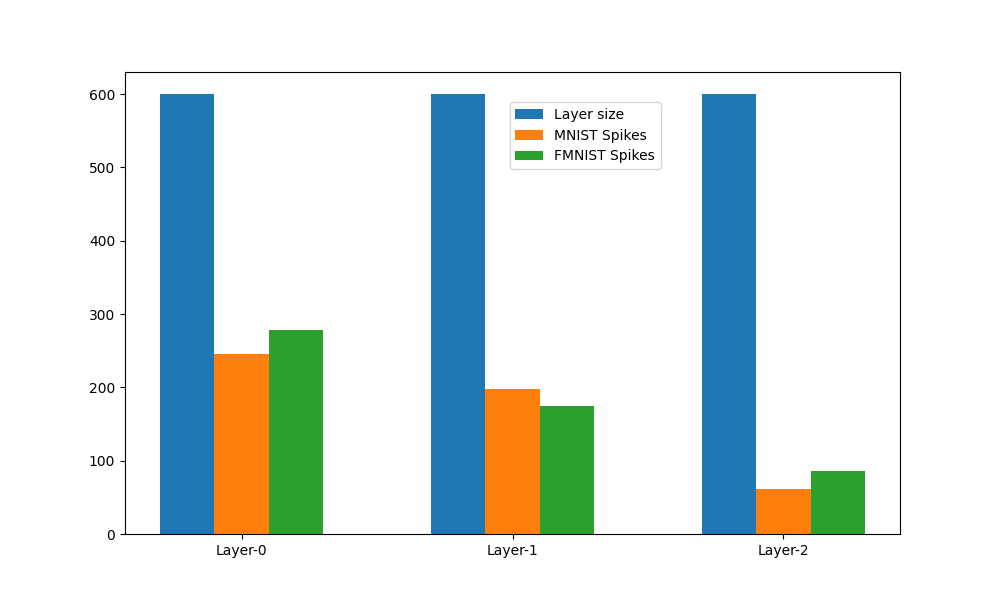}
		\vspace{-10pt}
		\caption{Ratio of firing neurons to layer size for a four-layer network model (784-600-600-600). The model uses population coding (detailed in Section \ref{sec:results}) The model's  accuracy is 96.2\% and 90.7\% for MNIST and FMNIST respectively}
		\label{fig:layerSize_vs_spikes}
		\vspace{-15pt}
\end{figure}

\section{Motivation for SNN DSE} \label{sec:methodology}
To motivate our approach, we start by studying the synaptic traffic or activity in the individual layers of the SNN. Section \ref{sec:exp_setup} details our experimental setup for this analysis. This analysis aims to find variabilities in the number of spiking neurons across layers. This fine-grained variability can be exploited to significantly improve the design of efficient hardware accelerators that satisfy application-specific latency, energy consumption, and area constraints. Figure \ref{fig:layerSize_vs_spikes} shows layer-wise variability using a fully-connected model with two hidden layers for the MNIST \cite{mnist} and FashionMNIST (FMNIST) \cite{fmnist} datasets. The model achieved 96.2\% accuracy for MNIST and 90.7\% accuracy for FMNIST. We used consistent layer sizes across the three hidden layers to monitor variabilities in the spiking activity independent of the number of neurons within each layer. 

Figure \ref{fig:layerSize_vs_spikes} shows that the number of firing neurons (averaged for five randomly selected time steps) declines exponentially as the layers get deeper. For example, in layer 0, the ratio of static neurons to firing neurons is 2.4. It increases to 3.4 and 10 for layer 1 and layer 2 respectively. We did not perform the analysis on deeper layers because they did not improve the accuracy of results for the datasets. \textbf{The key takeaway}: sparse firing traffic in deeper layers reduces the workload (i.e., accumulation of spikes) for post-synaptic layers. Consequently, this provides the opportunity to allocate fewer hardware neurons for those post-synaptic layers.

Deep networks might require prohibitive hardware resources for resource-constrained systems. However, based on a layer-wise variation analysis, resource allocation can be efficiently managed. As a result, DSE is imperative for evaluating the parameters of high-performing deep learning models. Hardware designers may also want to evaluate these models in comparison to each other (e.g., both ResNet and Lenet-5 perform with similar accuracy but ResNet occupies less hardware area) to enhance the design outcomes. Note that this experiment only shows spiking activity, but we also observed similar layer-wise variability for other model parameters, like weight quantization size, which significantly affects the system's memory requirements. Overall, an effective DSE approach will enable designers to explore the trade-off points of their SNN accelerator designs and provide feedback to their network models. This will result in a highly efficient hardware-software co-design process in terms of both model accuracy and hardware efficiency.


\begin{figure}[t]
\vspace{-15pt}
		\centering
		\includegraphics[width=0.99\linewidth]{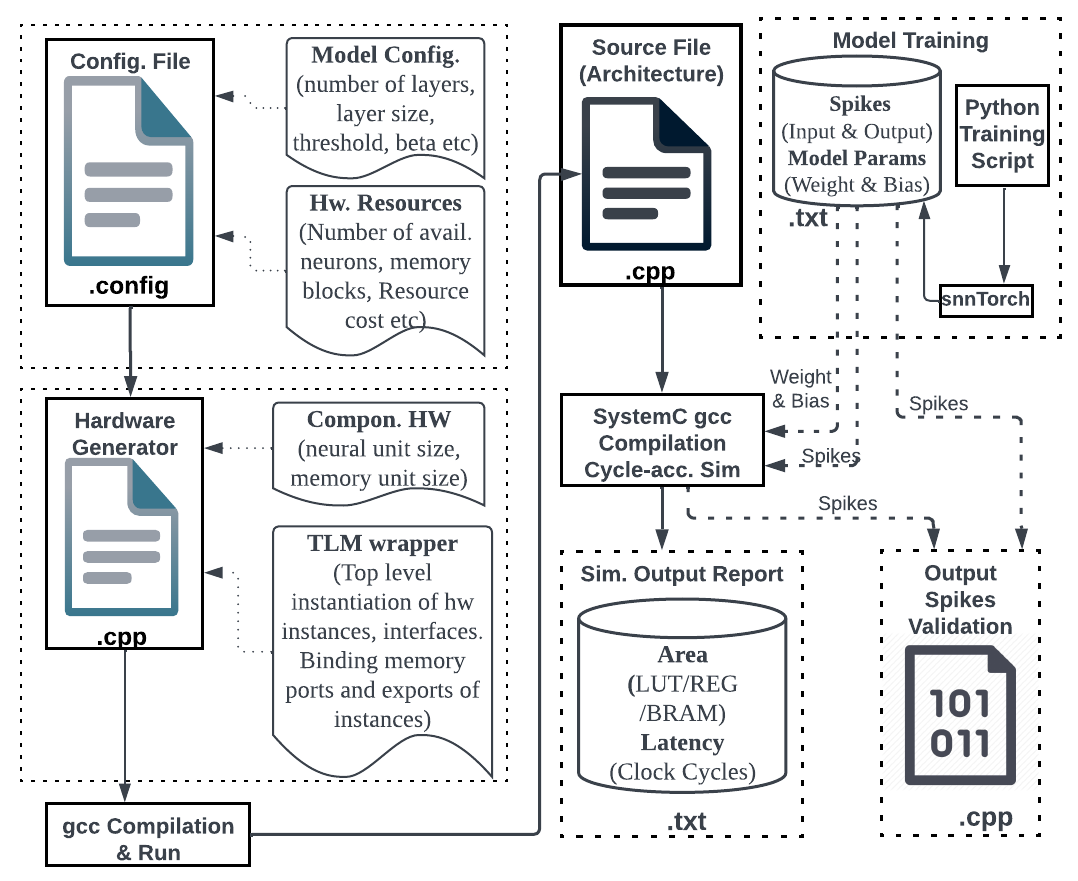}
		\vspace{-10pt}
		\caption{Overview of our framework outlining the key steps for rapid design space exploration}
		\label{fig:dse_method}
		\vspace{-15pt}
\end{figure}


\section{Design Space Exploration Methodology} \label{sec:methodology}
Figure \ref{fig:dse_method} depicts an overview of our framework and outlines the key functional components of the rapid DSE process. Since our work's main goal is to design efficient application-specific SNN accelerators, the starting point for an SNN DSE is a system specification that describes the network model for the target application. The following are the essential phases of the DSE methodology in our framework.

\noindent \textbf{Training Phase}: First, one or more candidate network topologies are selected and initially trained using a model simulation tool, like \textit{snntorch}\footnote{available online at \url{https://github.com/jeshraghian/snntorch}}. For clarity, we use \textit{snntorch} as a proxy for software machine learning libraries due to its native support for SNN simulations. Our framework includes a training script that orchestrates the training process given multiple models and selects the model that gives the best accuracy that is also within the desired accuracy range. It then extracts the input and output spikes and associated model parameters of the topology. Note that although the selection of the candidate topologies is mainly driven by the state-of-the-art network models (e.g., commonly used topologies for a certain dataset), we also experimented with models with random parameters to explore a larger design space toward a more accurate model. However, the initial network architecture search process is beyond the scope of this paper.
    
\noindent \textbf{Configuration Phase}: After training the target model and dumping its associated data, the data obtained from \textit{snntorch} is inserted into the configuration file (shown in the upper left corner in Figure \ref{fig:dse_method}). The model-related data include the number of hidden layers, the number of logical neurons in each corresponding layer, spike train length, and beta and threshold constants. In addition, the framework also sets the number of neurons per layer to define the logical-to-physical neuron ratio. This is an important hardware knob since realistic neural network models typically have too many neurons to be implemented or scaled in hardware. Moreover, unlike ANNs, SNN models naturally exhibit sparse spiking behavior, which leaves most of the neurons in an idle state. Our framework allows architectures (see Section \ref{sec:ssf_arch}) to exploit the sparsity of SNNs and explore this parameter in determining the mapping ratio. To enable an estimate of resource costs (e.g., lookup tables (LUT), registers, Block RAM (BRAM) primitives, etc.), our framework also features a library of hardware component costs that were obtained by synthesizing the individual hardware components. Additionally, the verbosity level of the simulation can be set for debugging and tracing purposes.
    
\noindent \textbf{Architecture Generation Phase}:  Next, the \textit{hardware generator} takes the configuration file and generates the corresponding detailed RTL architecture (bottom left corner in Figure \ref{fig:dse_method}). Adhering to the TLM guidelines, this script builds the target hardware architecture using the memory unit, neural unit, and event control from the hardware component libraries. In this process, the individual components are first modified to better suit each layer's model and hardware-specific constraints. For instance, the event control unit for hidden layer 0 will have a different state machine behavior than the other layers, depending on the total number of neurons. Similarly, the memory size will vary depending on the neural activity within each layer. We will describe the details of this architecture enhancement in Section \ref{sec:ssf_arch}. Given the component-level modifications, the framework also generates the top-level wrapper that couples the components together. In this process, it creates individual instances of the hardware components and connects their ports and exports.
    
\noindent \textbf{Simulation \& Validation Phase}: After generating the RTL architecture, the framework estimates the hardware resources for the target topology using the included component library (details in Section \ref{sec:exp_setup}). Then, it dumps the resource information into a text file that is used as input to a cycle-accurate SystemC simulation. At this stage, the simulator reads the model's input spikes along with the weight and bias data (from \textit{snntorch}) and simulates the inferred architecture. During simulation, it records the number of clock cycles as latency data for the SNN topology. Our framework also allows for the collection and recording of other peripheral execution data that might be useful for more detailed analysis. The data include the number and labels of spiking neurons in each layer and memory access counts. To verify the functionality of the generated architecture, the framework also performs a \textit{spike-to-spike validation} wherein the simulated output spikes are validated against the reference spikes from the trained input model.

\noindent \textbf{Evaluation Phase}: In this phase, both the model's performance (accuracy) and the hardware performance (latency and area cost) are evaluated. Depending on the evaluation result, modifications can be made to the hardware configuration (e.g., increase the neuron ratio, or reduce the memory blocks), after which further evaluation iterations would take place. Our framework can also automate the compilation and running of various configurations, which is a substantial advantage when the design space is large (this feature is omitted from Figure \ref{fig:dse_method} for brevity). Overall, utilizing a single Makefile, our framework is capable of conducting SNN DSE experiments with minimal user intervention, which would otherwise not be possible through RTL implementation.

\section{Implementation of the Framework}\label{sec:ssf_arch}\label{sec:architecture}

We implemented our hardware using SystemC \cite{panda2001systemc}, a C++ library for system-level modeling and hardware/software co-design. SystemC inherits all C++ features such as object-oriented programming (OOP) patterns and template-based meta-programming paradigms. These features are highly useful for defining an abstraction of a parametric processing element (PE) or any other hardware component in the TLM design (see Figure \ref{fig:tlm_platform}). In addition to C++ features, the SystemC library defines a set of enhanced features that makes it especially suitable for our work. For example, PE constructs can be modeled by \texttt{Module} entities of the SystemC library which is also inherited from \texttt{class} in the OOP. For PEs to communicate with each other, SystemC defines primitive channels and ports/exports (see Figure \ref{fig:tlm_platform}). Moreover, it also provides custom data types such as bit vectors (\texttt{sc\_bv}), arbitrary precision fixed point integers (\texttt{sc\_uint}), etc.

\subsection{Parametric Hardware Platform}

\begin{figure}[t!]
		\centering
		\includegraphics[width=1\linewidth]{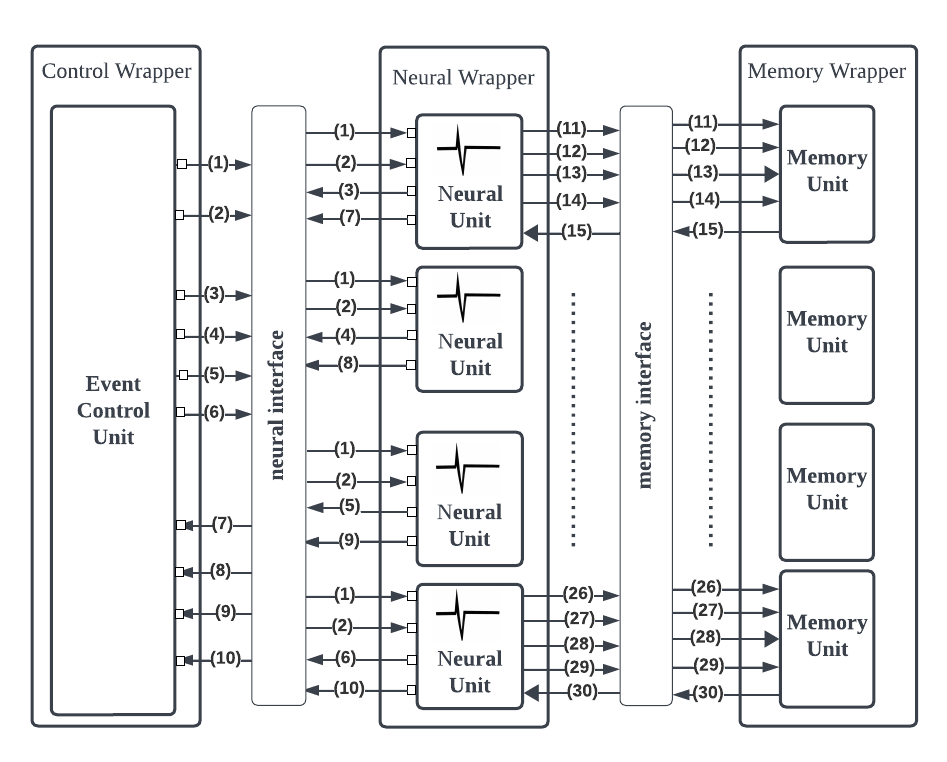}
		\vspace{-25pt}
		\caption{TLM-based Hardware Platform}
		\label{fig:tlm_platform}
		\vspace{-10pt}
\end{figure}

Figure \ref{fig:tlm_platform} depicts our generic TLM platform that represents a single layer of a network. Since our modeling utilizes the \textit{RTL/implementation-level} abstraction of TLM, wrappers are the parent class of the units that interact with the interface classes. These interface instances are the main ``communication channels'' through which computation components interact with each other. For all computation components, we use the clocked thread feature of SystemC to simulate cycle-accurate behavior. In this platform, a control wrapper and a neural wrapper form a single neural layer. Before moving on to the description of these basic components, we discuss our parallelization strategy for the SNN inference flow.

\noindent \textbf{Mapping Strategy}: The main challenge with a parallelization strategy is ability to keep hardware units always busy. For a fully Connected (FC) layer with $n$ neurons, our approach is straightforward: we partition the layer into $m$ groups (a design parameter): each group contains $n/m$ neurons, and each group is assigned to a Neural Unit (NU) during hardware synthesis. For example, in Figure \ref{fig:tlm_platform}, a layer is mapped to four neural units. For a Convolutional (CONV) layer, we parallelize output channel-wise, meaning that, for instance, each NU in Figure \ref{fig:tlm_platform} is responsible for $m$ output channels. Given this structure, we now define the processing flow of the spike trains. For this, we begin with a discussion of the Event Control Unit (ECU), which manages the spike-based processing flow. Note that the behavior of both FC and CONV ECUs are similar with minor distinctions.


\subsection{Event Control Unit} 
To provide a time-step-based processing flow, an ECU communicates with the pre- and post-synaptic layer ECUs to keep track of time steps and stay synchronized. Basically, it receives a spike train when the pre-synaptic layer has one ready. Likewise, it notifies the post-synaptic layer once its own spike train is ready. Intuitively, our simulator employs layer-wise pipelining: instead of having to wait for the post-synaptic layer, the ECU loads the spike train into a buffer and moves on to the next spike train from the pre-synaptic layer.

\begin{figure}[t!]
		\centering
		\includegraphics[width=1.0\linewidth]{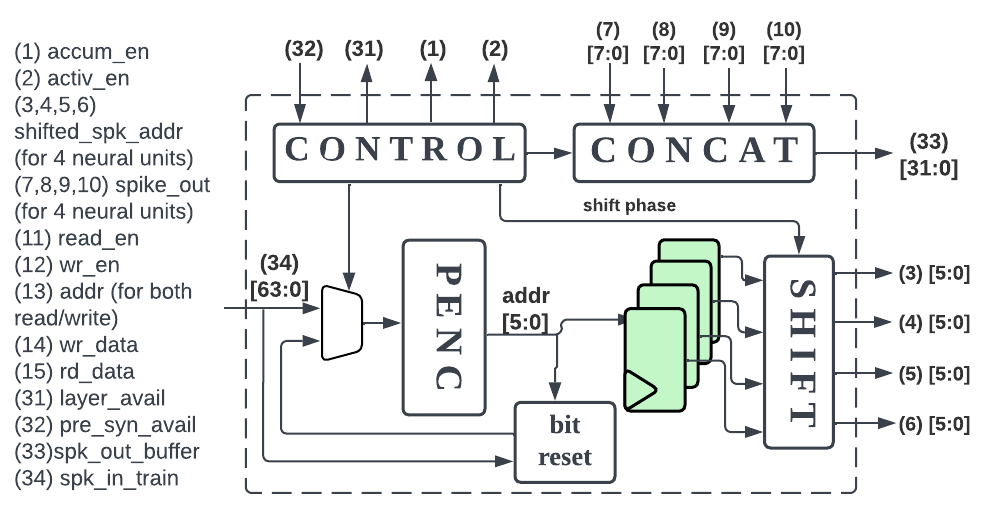}
		\vspace{-10pt}
		\caption{Event Control Unit (ECU) design}
		\label{fig:ecu}
		\vspace{-15pt}
\end{figure}

Within the ECU, a state machine orchestrates the spiking activity for the assigned neurons as depicted in Figure \ref{fig:ecu}. When it receives a spike train, it applies a compression mechanism to eliminate the non-spiking (e.g., reset) bits. With this mechanism, an $n$-bit spike train is translated into a shift register array (see Figure \ref{fig:ecu}). The process is as follows: in each cycle, the Priority Encoder (PENC) takes in $n$ bits of data and outputs the address of the first set bit, which gets written into the shift register array. The bit reset component of the ECU then resets the bit value of 1 for this address in the one cycle earlier version of the $n$-bit spike data. Despite inherent 2D $(row,col)$ nature of spikes in CONV layer, we store addresses in 1D fashion for the following reasons: (1) both PENC and Accumulation phases operate on 1D structure more efficiently, and (2) conversion between 1D and 2D is relatively lower cost in hardware, e.g., subtracting and adding (\ref{sec:neuralUnit} for details). From an FPGA hardware perspective, the PENC would ideally handle up to 100-bit inputs, beyond which the resource overhead would likely be prohibitive due to the FPGA routing overhead. Hence, PENC handles large inputs in chunks, meaning it compresses a subset of spikes to construct the general address set of the input spike train.

\subsection{Neural Unit} \label{sec:neuralUnit}
To provide fully-automated model mapping, we initialize each NU with a ``\textit{base address}" and ``\textit{neural size}" module parameters. In the context of an FC layer, this indicates that NU is responsible for logical neurons from (\textit{base address}) to (\textit{base address} + \textit{neural size}). The provided shift address also serves as the weight address for the synapse memory. The NU iterates through its neurons and serially calculates their accumulator values. Once the ECU transitions from the accumulation to the activation phase, using the Leaky Integrate and Fire (LIF) neuron model, the NU calculates the membrane potential for the neurons. For this, it adds three components together: (i) leaky potential value (multiplied potential from the previous time step with the beta constant), (ii) the accumulated value from the shifting phase, and (iii) the neuron bias. Then, the NU checks whether the new membrane value exceeds the threshold and assigns a spike based on the result.

\begin{figure}[t!]
		\centering
		\includegraphics[width=0.7\linewidth]{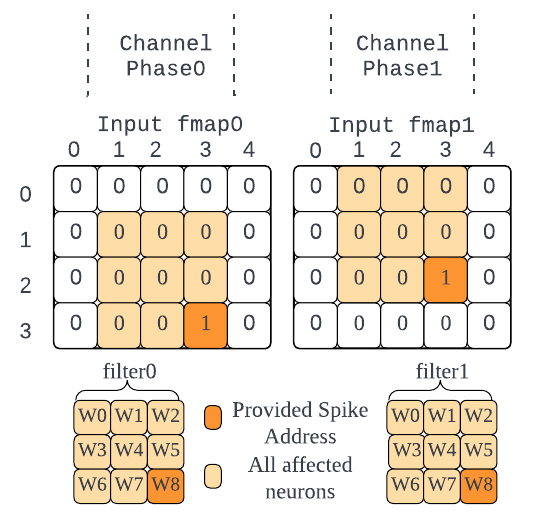}
		\vspace{-10pt}
		\caption{The illustration for spike to neuron address extraction and weight accumulation flow.}
		\label{fig:conv_illustr}
		\vspace{-15pt}
\end{figure}

In the case of the CONV layer, the NU is responsible for the output channels ranging from (\textit{base address}) to (\textit{base address} + \textit{neural size}). For each output channel assigned, an NU serially processes spikes from each input feature map (fmap). \cite{spike_conv_propose_2018} proposed the spike-based convolution first. As figure \ref{fig:conv_illustr} illustrates, for a given input spike address, the NU calculates the addresses for all affected neurons, which also depends on the filter size (a design parameter). For the filter size of three, there are nine neurons impacted by this spike (unless the addresses do not exceed the frame). The NU serially reads the membrane potential values for the affected neuron addresses and adds the corresponding filter coefficients to the potential values. Note that \cite{spike_conv_propose_2018} employs input channel-wise parallelization: a spike from each input fmap is processed in parallel, whereas output channel-wise in our design. Therefore, the NU serially iterates through all input channels, and then it performs the activation/spiking operation. Finally, to implement max-pooling in hardware, we OR-gate the generated spike train with $2x2$ window size in non-overlapping fashion \cite{event_conv_2022}.



\subsection{Memory Unit} This unit has memory blocks that store synapse weight information and mapping logic that manages the access of multiple hardware neurons to a single memory block. 
Our platform lets users set the depth and count of the memory blocks. The depth of the blocks can be configured to $M \times SIZE$ where $M$ is the number of neurons assigned per memory block and $SIZE$ is the size of the pre-synaptic layer. As discussed in Section \ref{sec:neuralUnit}, the memory unit uses the \textit{memory interface} to respond to the weight read requests. 

\textbf{Memory Interface}: The communication between the neural unit and  memory unit is established via the \textit{memory interface} class which is a virtual class whose behavior is purely implemented in the calling class. In TLM terminology, the calling class is named the \textit{export} class since information is being exported into it. Therefore, the implementation of the interface class consists of a set of methods to be invoked by the export class. In this platform, the memory interface has a \textit{Read} method for reading a specific synapse weight. The method utilizes two signals (e.g., signal labels 16 and 17 in Figure \ref{fig:tlm_platform}): a 32-bit \texttt{read\_data} bus that carries weight information and a 1-bit \texttt{read\_en} line to enable data reads.

\textbf{Neural Interface}: The neural interface for the event control unit is more sophisticated than the memory interface. It contains both \textit{Read} and \textit{Write} methods to communicate the following signals (e.g., signal labels 1 to 8 in Figure \ref{fig:tlm_platform}): \texttt{accumulation\_en} (1-bit) and \texttt{activation\_en} (1-bit) are the enable signals used to allow all neurons within the neural unit to perform accumulation and activation operations. \texttt{shifted\_spike\_addr} (\texttt{N}-bit) represents the address to the individual neurons from the pre-synaptic layer whose weight is to be integrated, where \texttt{N} is the size of pre-synaptic layer. The \texttt{spike\_out} shows whether the neuron spiked or not and \texttt{done} signals that its associated neuron completed accumulation or activation. Note that the number of buses for the last three signals depends on the number of neurons in the neural unit, which can be specified by the user in the configuration file.

\begin{table*}[t]
\begin{threeparttable}
\centering
    \caption{Summary of results for sparsity-aware neuron resource allocation under different layer-wise LHR parameter sets. \textbf{TW} denotes This Work. \textbf{LUT - Lat. Impr.} stands for improvement/reduction in LUT and latency compared to the baseline (prior work). \textbf{Pop. Cod.} refers to the size of population neurons in the output layer. Average Spike Events per layer per network for net-1 \(784(95)-500(81)-500(86)-300\), net-2 \(784(118)-300(98)-300(56)-200\), net-3 \(784(186)-1024(321)-1024(304)-300\), net-4 \(784(316)-512(169)-256(87)-128(37)-64(20)-150\),
    and net-5 (128$\times$128(135)-32C3(240)-P2-32C3(1250)-P2-512(21)-256}
    \vspace{-15pt}
    \begin{tabular}{|c||c|c|c|c|c|c|c|c|c|}
        \hline
         & \textbf{Net.} & & \textbf{Target} & \textbf{Est. Area} & \textbf{Cycles/} & \textbf{LUT - Lat.} & \textbf{Energy/} &\textbf{Pop.} & \textbf{Acc.} \\
        \textbf{Dataset} & \textbf{Top.} & \textbf{Work} & \textbf{Device} & \textbf{LUT/REG} & \textbf{Cod.} & \textbf{Impr.} & \textbf{Image} & \textbf{Image}  & [\%] \\
            \hline \hline
            \multirow{14}{*}{MNIST} 
            & \multirow{5}{*}{(net-1)}    
            & \cite{fang2020encoding} & Zynq US+ & 124.6K/185.2K & 65,000 & --- & 2.34 mJ & --- & 98.96\\
            \cline{3-10}
            &\multirow{5}{*}{784-500-}& \textbf{TW}${-(1, 1, 1)}$ && 157.6K/103.1K & 10,583 & $\times$1.26, $\times$0.16 & 0.09 mJ & & \\
            &\multirow{5}{*}{500-10}& \textbf{TW}${-(2, 1, 1)}$ & \multirow{3}{*}{Virtex US+} & 127.2K/83.2K & 16,807 &$\times$1.02, $\times$0.26 & 0.12 mJ & \multirow{4}{*}{300} & \multirow{4}{*}{97.52} \\
            && \textbf{TW}${-(1, 2, 1)}$ && 127.2K/83.2K & 15,561 & $\times$1.02, $\times$0.24 & 0.11 mJ & & \\
            && \textbf{TW}${-(4, 4, 4)}$ && 60.8K/39.7K & 31,583 & $\times$0.49, $\times$0.48 & 0.17 mJ & & \\      
             && \textbf{TW}${-(4, 8, 8)}$ && 30.7K/63.4K & 53,308 & $\times$0.24, $\times$0.82 & 0.27 mJ& & \\            
            \cline{2-10}

            & \multirow{4}{*}{(net-2)}    
            & \cite{abderrahmane2020design} & Cyclone V & 22.8K/9.3K & 1,660 & --- & --- & --- & 98.96\\
            \cline{3-10}
            &\multirow{4}{*}{784-300-} & \textbf{TW}${-(1, 1, 1, 1)}$ & & 136.5K/86.1K & 18,710 & $\times6$, $\times0.05$ & 0.14 mJ & & \\
            &\multirow{4}{*}{300-300-}& \textbf{TW}${-(4, 4, 4, 1)}$ & \multirow{3}{*}{Virtex US+} & 54.9K/33.2K & 67,586 &$\times2.33$, $\times0.19$ & 0.39 mJ & \multirow{4}{*}{200} & \multirow{4}{*}{98.02} \\

            &\multirow{4}{*}{10}& \textbf{TW}${-(4, 4, 8, 1)}$ & & 50.5K/30.2K & 68,542 & $\times2.11$, $\times0.19$ & 0.39 mJ & & \\
            && \textbf{TW}${-(2, 2, 16, 8)}$ && 45.7K/27.2K & 69,998 & $\times0.88$, $\times0.2$ & 0.37 mJ & & \\  
            && \textbf{TW}${-(4, 4, 16, 8)}$ && 27.5K/15.4K & 72,330 & $\times1.05$, $\times0.21$ & 0.36 mJ & & \\ 
            \hline

            \multirow{13}{*}{FMNIST}
            & \multirow{5}{*}{(net-3)}    
            & \cite{FPGANHAP} & Kintex-7 & 124.6K/185.2K & 65,000 & --- & 2.23 mJ  & --- & 86.97\\
            \cline{3-10}
            &\multirow{5}{*}{784-1024-}& \textbf{TW}${-(1, 1, 1)}$ && 287.6K/185.5K & 34,563 & $\times23.54$, $\times0.02$ & 1.12 mJ & & \\
            &\multirow{5}{*}{1024-10}& \textbf{TW}${-(2, 1, 1)}$ & \multirow{3}{*}{Virtex US+} & 225.7K/145.2K & 35,011 &$\times18.47$, $\times0.02$ & 0.97 mJ & \multirow{4}{*}{300} & \multirow{4}{*}{84.41} \\
            && \textbf{TW}${-(8, 2, 4)}$ && 90.8K/56.2K & 96,827 & $\times7.37$, $\times0.06$ & 1.37 mJ & & \\
             && \textbf{TW}${-(16, 8, 4)}$ && 35.8K/21.4K & 187,099 & $\times2.93$, $\times0.11$ & 1.45 mJ & & \\  
            && \textbf{TW}${-(32, 32, 8)}$ && 13.9K/8.7K & 388,897 & $\times1.13$, $\times$0.24 & 2.21 mJ & & \\             
            \cline{2-10}

            & \multirow{4}{*}{(net-4)}    
            & \multirow{2}{*}{\cite{ye2022implementation}} & \multirow{2}{*}{Kintex-7} & \multirow{2}{*}{13.7K/12.4K} & \multirow{2}{*}{1,562K} & \multirow{2}{*}{---} & \multirow{2}{*}{---} & \multirow{2}{*}{---} & 48.92\\
            \cline{10-10}
            &&&&&&&&&85.38 \\
            \cline{3-10}
            &\multirow{4}{*}{784-512-} & \textbf{TW}${-(1, 1, 1, 1, 1)}$ && 137.8K/90.3K & 40,142 & $\times10.01$, $\times0.00$ & 0.56 mJ & & \\
            &\multirow{4}{*}{256-128-}& \textbf{TW}${-(1, 4, 4, 1, 1)}$ & \multirow{3}{*}{Virtex US+} & 103.1K/69.8K & 61,724 &$\times$7.48, $\times0.00$ & 0.73 mJ & \multirow{4}{*}{150} & \multirow{4}{*}{76.4} \\
            &\multirow{4}{*}{64-10}& \textbf{TW}${-(2, 8, 4, 16, 8)}$ && 45.1K/67.2K & 114,266 & $\times3.27$, $\times0.00$ & 0.9 mJ & & \\
            && \textbf{TW}${-(4, 2, 8, 8, 64)}$ && 37.7K/24.6K & 69,534 & $\times2.74$, $\times0.00$ & 0.48 mJ & & \\ 
            && \textbf{TW}${-(32, 16, 8, 16, 64)}$ && 6.6K/63.4K & 843,518 & $\times0.73$, $\times0.03$ & 4.3 mJ & & \\ 
            \hline

            \multirow{5}{*}{DVS} 
            & \multirow{1}{*}{(net-5)}    
            & \cite{di2022sne_dvsgest} & 22nm ASIC & -- & 6,044K & --- & 0.17 mJ & --- & 92.42\\
            \cline{3-10}
            \multirow{5}{*}{128} &\multirow{1}{*}{128$\times$128-}& \textbf{TW}${-(1, 1, 8, 32)}$ && 137.5K/361.5K & 2,481K & --, $\times$0.41 & 14.93 mJ & ---  & \\
            \multirow{5}{*}{Gesture}&\multirow{1}{*}{32C3-P2-}& \textbf{TW}${-(1, 1, 16, 16)}$ & \multirow{3}{*}{Virtex US+} & 128.1K/352.1K & 2,493K &--, $\times$0.41  & 13.41 mJ & \multirow{4}{*}{--} & \multirow{4}{*}{71.23} \\
            &\multirow{1}{*}{32C3-P2-}& \textbf{TW}${-(1, 1, 32, 32)}$ && 119.2K/343.7K & 4,475K & --, $\times$0.74 & 20.5 mJ & & \\
            &\multirow{1}{*}{512-256-}& \textbf{TW}${-(1, 1, 16, 256)}$ && 123.4K/347.5K & 2,521K & --, $\times$0.40 &7.21 mJ &  & \\      
             &\multirow{1}{*}{11}& \textbf{TW}${-(16, 1, 16, 256)}$ && 93.5K/267.5K & 2,486K & --, $\times$0.41 & 6.24 mJ &  & \\  
            \hline
    \end{tabular}
    \label{tab:res_alloc}
    \end{threeparttable}
 \end{table*}

\section{Experimental Results} \label{sec:results}

\subsection{Experimental Setup}\label{sec:exp_setup}
We use C++ and SystemC 2.0 to implement the framework’s software (which simulates the SNN hardware). The hardware components are developed in SystemVerilog RTL and the generated hardware instances were synthesized using Xilinx Vivado onto a Xilinx Virtex® UltraScale+™ FPGA with a 100MHz clock frequency to obtain precise FPGA area reports. We provide resource utilization results in Table \ref{tab:res_alloc}.
As we mentioned in Section \ref{sec:ssf_arch}, we utilize the \textit{snntorch} library for training. Within \textit{snntorch}, two primary methods are typically employed: Surrogate Gradient Descent (SGD) and Backpropagation Through Time (BPTT) \cite{li2021differentiable}. SGD is a technique that addresses the non-differentiable nature of the spiking mechanism in SNNs. It substitutes the original non-differentiable function with a smooth surrogate derivative that allows the usage of conventional gradient descent methods for optimization. On the other hand, BPTT is a temporal variant of the traditional backpropagation algorithm, which considers the recurrent nature of SNNs. We employ SGD for our models as it captures precise spike timings.

We use the static (MNIST and FMNIST) and dynamic (DVSGesture) datasets as driving applications. The static datasets contain 28$\times$28 grayscale image samples. DVSGesture contains 128$\times$128 frames for hand gesture recognition. Each frame captures changes in pixel intensity using Dynamic Vision Sensor cameras. To evaluate our framework, we compare it with existing state-of-the-art SNN inference accelerators, as there are no dedicated simulators for SNN hardware. We rigorously evaluate our framework's latency in clock cycles and compare it to the results of five prior SNN accelerators \cite{fang2020encoding,abderrahmane2020design,FPGANHAP,ye2022implementation, di2022sne_dvsgest}. The second column in Table \ref{tab:res_alloc} summarizes the SNN model topologies for which these previous accelerators were designed. Net-1 to net-4 are fully connected (FC) networks with different numbers of hidden layers. Net-5 is 32C3-P2-32C3-P2-512-256-11 where 32C3 stands for 32 filters with size of $3\times3$ and P2 for maxpooling with size of $2\times2$  followed by three fully connected layers.

\subsection{Impact of Logical-to-Hardware Neuron Ratio}
Table \ref{tab:res_alloc} shows how different layer-wise logical-to-hardware ratios ($LHR$) affect the latency per inference and the resource utilization of our flexible hardware design. 
$LHR$ is a parameter that controls the mapping ratio of model hyperparameters into hardware. For fully connected layers, $LHR$ indicates the number of logical neurons per physical hardware neuron (i.e., Neural Unit) in each layer of the network. For convolutional layers, $LHR$ indicates the number of logical output channels per Neural Unit. For example, ($LHR - 1, 2, 4, 1$) for net-5 means that the network has four hidden layers. Each neural unit in the first and second layers handles one and two output channels, respectively, and each neural unit handles four logical neurons in the third layer and one neuron in the fourth layer. See Section \ref{sec:ssf_arch} for more details.

We use LUT-Latency improvement (depicted as \textbf{LUT-Latency Impr.} in Table \ref{tab:res_alloc}) as a metric to measure the improvement in FPGA area (LUT) and inference latency (i.e., total clock cycles for inferring a single test sample) over the prior works.

We vary $LHR$ for each layer (by powers of two) for each network topology to explore the trade-offs between LUT and latency across datasets and topologies. In some cases, our baseline design may have worse latency or LUT than the prior works, mainly because the prior works are optimized for their specific fixed hardware configurations. However, by tuning $LHR$, we can achieve similar or better efficiency in terms of either latency or LUT. For example, for the MNIST dataset, our design with ${(LHR - 4, 8, 8)}$ for topology (1) reduces LUT by $76\%$ and maintains the same latency as \cite{fang2020encoding}, and our design with ${(LHR - 4, 4, 16, 8)}$ for topology (2) achieves 0.21× latency with similar LUT as \cite{abderrahmane2020design}. Note that Fang et al. \cite{fang2020encoding} do not report the PE size (which determines the parallelism of neuronal operations) for their synthesis results, although they claim that their PE size is parametric. On the other hand, Abderrahman et al. \cite{abderrahmane2020design} state that their design executes the first hidden layer in fully-parallel mode and the rest of the layers in serial mode with only one hardware neuron per layer. Similarly, for the Fashion MNIST dataset, our design with ${(LHR - 32, 32, 8)}$ for topology (3) outperforms the baseline by $4.1\times$ at the expense of 13\% more hardware resources compared to \cite{FPGANHAP} and ${(LHR - 32, 16, 8, 16, 64)}$ scheme for topology (4) outperforms the prior work by $31.25\times$ with 27\% less hardware than \cite{ye2022implementation}. Unlike latency, which scales more steeply, energy serves as a more balanced metric that takes both latency and area into consideration. Additionally, it's worth mentioning that in a fully realized hardware implementation, after area optimization, energy efficiency can be further enhanced through clock gating.

\begin{figure}[t!]
\vspace{-15pt}
		\centering
		\includegraphics[width=1.1\linewidth]{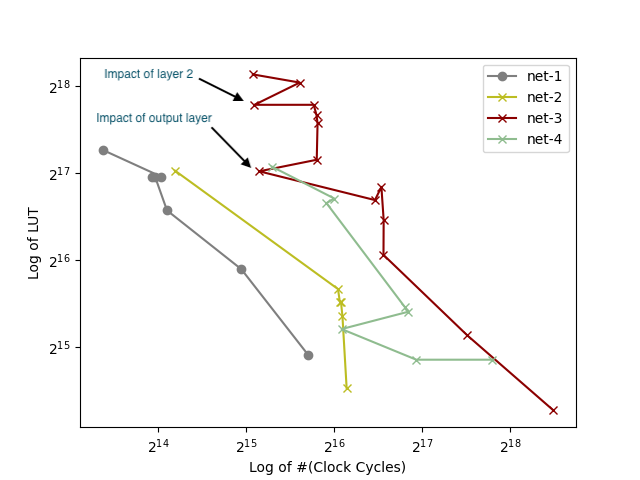}
		\vspace{-5pt}
		\caption{Overview of Latency-LUT trend for the topologies tabulated in Table \ref{tab:res_alloc}. Although clock cycles increase as the area decreases, in some cases the same (or even less) resource with different $LHR$ combinations leads to lower clock cycles.}
		\label{fig:lat-lut_trend}
		\vspace{-10pt}
\end{figure}

Our baseline (highest resource allocated) mapping scheme for DVSGesture performs 2.5$\times$ better in terms of cycles but 87.6$\times$ higher energy compared to a prior ASIC implementation \cite{di2022sne_dvsgest} (which is also sparsity-aware). Despite the high sparsity characteristic of the data (see Table \ref{tab:res_alloc} caption), the long latency can be attributed to the lengthy time steps required to achieve close to state-of-the-art (SoA) accuracy in snntorch. Yet, the highest attainable accuracy was $71.23\%$ with 124 time steps (and beta set to 0.23). In comparison, the prior work manages to achieve higher accuracy while applying maxpooling to the input layer, directly reducing the input frame size from 128 down to 32. We were unable to apply maxpooling due to low accuracy. Furthermore, while an ASIC implementation might offer significant energy advantages due to its tailored design, our approach provides a valuable balance of performance improvement, combined with the benefits of flexibility inherent to FPGA-based implementations.

Our layer-wise analysis of the network showed that the majority of processing time is consumed by the second convolutional layer followed by the first fully connected layer, which also has a high input spike activity as shown in Table \ref{tab:res_alloc} caption. Therefore, in the configurations $(LHR - 1, 1, 8, 32)$, $(LHR - 1, 1, 16, 16)$, $(LHR - 16, 1, 16, 256)$, latency remains consistent largely because the second convolutional layer alone overshadows other layers' latencies in the pipeline. For $(LHR - 1, 1, 32, 32)$, latency increases due to the increased workload in the first fully connected layer's neural unit. Based on this analysis, we conclude that the $(LHR - 16, 1, 16, 256)$ configuration is the best mapping for this use case due to the reduction in the hardware area, which translates into lower inference energy. Importantly, our approach enables rapid exploration of the design space to achieve a 64\% reduction in the inference energy compared to the sparsity-oblivious baseline scheme, while maintaining the same latency.

Figure \ref{fig:lat-lut_trend} captures the high-level view of the Latency-LUT trend for the same topologies as in Table \ref{tab:res_alloc}. Some trends have irregular patterns (i.e., lower latency despite reducing LUT) because of the layer-wise allocation of hardware neurons. For instance, in net-3, hidden layer 1 and hidden layer 2  have the highest spike events (see Table \ref{tab:res_alloc} footnotes) and hence dominate the network latency. Therefore, a slight reduction in resources leads to a significant performance degradation for the network. In general, we observed that the spike event counts in Table \ref{tab:res_alloc} follow a ratio of $1/3$ of the layer size for the first layer and about $2/7$ for the second hidden layer. This is consistent with existing works \cite{yin2022sata} that suggest that the sparsity increases as the network gets deeper.

\subsection{Spike Train Length vs. Population Coding Ratio}

\begin{figure*}[t]
  \centering
  \subfloat[]{
    \includegraphics[width=0.49\textwidth]{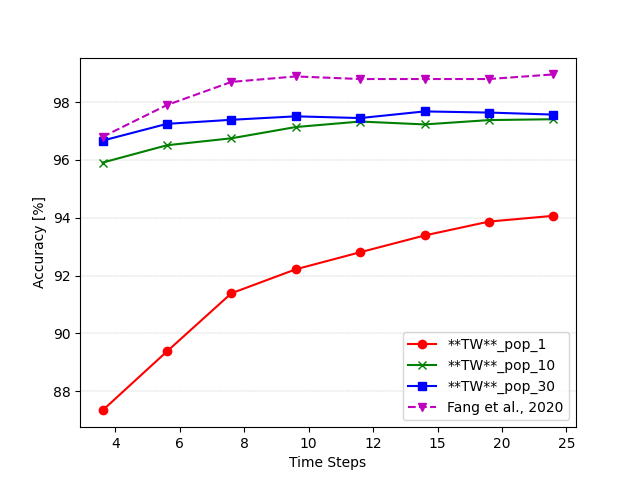}
    \label{fig:accuracy}
  }
  \subfloat[]{
    \includegraphics[width=0.44\textwidth]{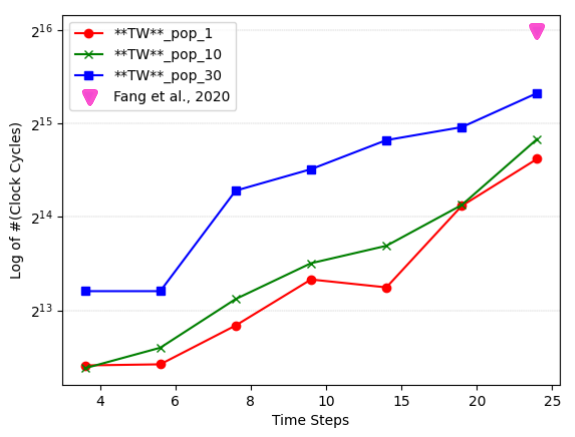}
    \label{fig:latency}
  }
  \vspace{-5pt}
    \caption{Spike train length sweep: scaling of (a) accuracy and (b) latency (clock cycles) for MNIST with different population coding ratios. $Pop\_1$, $\_10$, and $\_30$ denote population ratios of 1, 10, and 30, respectively. The figures illustrate the trade-offs between population coding ratios (i.e., number of output neurons) and spike train length of this work (TW) compared to prior work \cite{fang2020encoding}}.
    \label{fig:spike_train_len}
    \vspace{-15pt}
\end{figure*}

A major, yet under-explored, hyperparameter in the SNN design space is spike train length. The spike train length specifies the length of the encoding window required to transform real-valued images (pixel resolution) into spikes. In general, a short spike train length leads to poor accuracy but fast computation time due to low precision during conversion and inadequate time for the neuron to complete the accumulation, e.g., there are not enough time steps to produce spikes. This drawback can be mitigated by employing a coding scheme known as ``population coding" over the output layer of the network \cite{temp_vs_spatial_coding}. Indeed, a study published in \textit{Current Opinion in Neurobiology} \cite{pop_coding_brain} has shown that the brain extensively employs population coding in certain regions for efficient information representation. With this coding scheme applied to the SNN's classification layer, each class or category is represented by a pool of neurons, e.g., 10 neurons per class of the 10 categories in the MNIST dataset. Hence, we define population coding ratio (PCR) as a parameter that controls how many logical neurons are assigned per class.

We investigate the combined impact of spike encoding window length and PCR on model accuracy and hardware latency. We vary the spike train length from four to 25 time steps in Figure \ref{fig:spike_train_len} with three different PCRs (\textit{TW\_pop\_1} for one neuron per class, \textit{TW\_pop\_10} for 10 neurons per class, \textit{TW\_pop\_30} for 30 neurons per class) and show the scaling of the model's (a) accuracy vs. (b) latency (in clock cycles) for an MNIST image. We also compare our results with a previous work that performed a similar experiment on the same network topology (784-500-500-10). As the spike train length is increased, we observe a significant improvement in performance for \textit{TW\_pop\_1} as the spike train length increase from four to 20, as shown in Figure \ref{fig:accuracy}; we observe no improvement beyond 25 (i.e., the best attainable accuracy at 25 is 94.07\%). In contrast, we notice the immediate effect of population coding in \textit{TW\_pop\_10} and \textit{TW\_pop\_30} where the accuracy starts at 96\%, even for short spike train length, and continues to slightly increase. For \textit{TW\_pop\_30}, we achieve 97.68\% accuracy at time steps 15 after which we observe a slight drop due to potential model over-fitting. Therefore, we can conclude that 15 time steps are sufficient to represent MNIST images with high resolution although most of the existing works use 50 or more time steps \cite{lee2016training}.

In comparison, Fang et al. \cite{fang2020encoding} outperforms our accuracy by 1.85\% and achieves the highest accuracy of 98.96\% at time steps 25. The superiority of this prior work can be attributed to the optimized spike encoding schemes as opposed to the standard rate coding utilized in this work. In terms of latency, however, our clock cycle results for the best accuracy (29008) outperform the prior work by more than $2\times$ as shown in Figure \ref{fig:latency}. While both the prior work and ours employ similar hardware design strategies and execute in a layer-wise pipelined manner, the latency savings can be attributed to their PE size (e.g., hardware neurons) which, as noted before, is not disclosed in their discussion.

Higher PCR ratios lead to longer latency since more shifting iterations are required to propagate spikes from the pre-synaptic neurons to the output layer. For instance, latency leaps by $2\times$ for \textit{TW\_pop\_30} when we change time steps from six to eight, whereas we observe a substantially lower pace of scaling in \textit{TW\_pop\_10} by $1.43\times$. Another major drawback of the neural population coding is the increase in total neuron count. However, we argue that our design and the design space exploration enabled by our work can help to mitigate both drawbacks. In terms of latency, the design executes in a layer-wise pipelined manner. Moreover, the output layer is typically the smallest across the network and is inherently highly sparse (i.e., a lower number of spike events). Hence the increased execution cycles in the output layer do not directly translate into overall latency since that time would otherwise be spent by stalling the layer while waiting for the next spike train from the pre-synaptic layer. Overall, the key goal of this experiment is to demonstrate the ability of population coding to project temporal information into the spatial domain, thus favoring the inference latency of the network.

\section{Conclusion and Future Work}

This article presented and demonstrated the effectiveness of sparsity-aware design space exploration for SNN hardware accelerators. Specifically, we have shown the benefits of utilizing layer-wise sparsity in SNNs, which we argue is a grand challenge for SNNs and a crucial consideration toward achieving brain-like hardware efficiency. We present a sparsity-driven hardware neuron allocation approach that can achieve up to $76\%$ savings in hardware resources while maintaining a similar latency to prior SNN accelerators that do not consider sparsity. We also investigated the effects of two important model hyperparameters---spike train length and neuron population size---on SNN acceleration. Both hyperparameters have a significant impact on the trade-off between hardware performance and model accuracy. We further showed that the population coding technique is particularly advantageous for our design compared to previous work, since sparsity occurs least at the output layer thereby leading to minimal hardware overhead for our design. 

For future research, we aim to implement a dynamic (runtime) scheme of sparsity-aware neuron allocation directly in hardware and explore the deployment of FPGA-based SNN accelerators for a wider variety of SNN models and datasets. Moreover, we plan to conduct detailed comparative analyses of SNNs against ANNs with a heavy focus on sparsity. We aim to delve deeply into the question of how to exploit the inherent potential efficiency benefits of SNNs, characterized by their simpler computations, to maintain a competitive edge over traditional ANNs in terms of computational efficiency.

\bibliographystyle{ieeetr}
{\small
\bibliography{refs}

\begin{thebibliography}{10}

\bibitem{rathi2022exploring}
N.~Rathi, I.~Chakraborty, A.~Kosta, A.~Sengupta, A.~Ankit, P.~Panda, and
  K.~Roy, ``Exploring neuromorphic computing based on spiking neural networks:
  Algorithms to hardware,'' {\em ACM Computing Surveys}, 2022.

\bibitem{foldiak2003sparse}
P.~Foldiak, ``Sparse coding in the primate cortex,'' {\em The handbook of brain
  theory and neural networks}, 2003.

\bibitem{faghihi2022neuroscience}
F.~Faghihi, S.~Cai, and A.~A. Moustafa, ``A neuroscience-inspired spiking
  neural network for eeg-based auditory spatial attention detection,'' {\em
  Neural Networks}, vol.~152, pp.~555--565, 2022.

\bibitem{truenorth}
F.~Akopyan, J.~Sawada, A.~Cassidy, R.~Alvarez-Icaza, J.~Arthur, P.~Merolla,
  N.~Imam, Y.~Nakamura, P.~Datta, G.-J. Nam, {\em et~al.}, ``Truenorth: Design
  and tool flow of a 65 mw 1 million neuron programmable neurosynaptic chip,''
  {\em IEEE transactions on computer-aided design of integrated circuits and
  systems}, vol.~34, no.~10, pp.~1537--1557, 2015.

\bibitem{davies2018loihi}
M.~Davies, N.~Srinivasa, T.-H. Lin, G.~Chinya, Y.~Cao, S.~H. Choday, G.~Dimou,
  P.~Joshi, N.~Imam, S.~Jain, {\em et~al.}, ``Loihi: A neuromorphic manycore
  processor with on-chip learning,'' {\em Ieee Micro}, vol.~38, no.~1,
  pp.~82--99, 2018.

\bibitem{furber2014spinnnaker}
S.~Furber, ``Spinnnaker: The world's biggest noc,'' in {\em 2014 Eighth
  IEEE/ACM International Symposium on Networks-on-Chip (NoCS)}, pp.~ii--ii,
  IEEE, 2014.

\bibitem{neil2014minitaur}
D.~Neil and S.-C. Liu, ``Minitaur, an event-driven fpga-based spiking network
  accelerator,'' {\em IEEE Transactions on Very Large Scale Integration (VLSI)
  Systems}, vol.~22, no.~12, pp.~2621--2628, 2014.

\bibitem{khodamoradi2021s2n2}
A.~Khodamoradi, K.~Denolf, and R.~Kastner, ``S2n2: A fpga accelerator for
  streaming spiking neural networks,'' in {\em The 2021 ACM/SIGDA International
  Symposium on Field-Programmable Gate Arrays}, pp.~194--205, 2021.

\bibitem{li2022efficiency}
Z.~Li, E.~Lemaire, N.~Abderrahmane, S.~Bilavarn, and B.~Miramond, ``Efficiency
  analysis of artificial vs. spiking neural networks on fpgas,'' {\em Journal
  of Systems Architecture}, vol.~133, p.~102765, 2022.

\bibitem{yin2022sata}
R.~Yin, A.~Moitra, A.~Bhattacharjee, Y.~Kim, and P.~Panda, ``Sata:
  Sparsity-aware training accelerator for spiking neural networks,'' {\em IEEE
  Transactions on Computer-Aided Design of Integrated Circuits and Systems},
  2022.

\bibitem{abderrahmane2020design}
N.~Abderrahmane, E.~Lemaire, and B.~Miramond, ``Design space exploration of
  hardware spiking neurons for embedded artificial intelligence,'' {\em Neural
  Networks}, vol.~121, pp.~366--386, 2020.

\bibitem{fang2020encoding}
H.~Fang, Z.~Mei, A.~Shrestha, Z.~Zhao, Y.~Li, and Q.~Qiu, ``Encoding, model,
  and architecture: Systematic optimization for spiking neural network in
  fpgas,'' in {\em 2020 IEEE/ACM International Conference On Computer Aided
  Design (ICCAD)}, pp.~1--9, IEEE, 2020.

\bibitem{2022_hw_footprint}
E.~Lemaire, B.~Miramond, S.~Bilavarn, H.~Saoud, and N.~Abderrahmane, ``Synaptic
  activity and hardware footprint of spiking neural networks in digital
  neuromorphic systems,'' {\em ACM Transactions on Embedded Computing Systems},
  vol.~21, no.~6, pp.~1--26, 2022.

\bibitem{SIS_2023_CAS}
A.~J. Leigh, M.~Heidarpur, and M.~Mirhassani, ``Digital hardware
  implementations of spiking neural networks with selective input sparsity for
  edge inferences in controlled image acquisition environments,'' {\em IEEE
  Transactions on Circuits and Systems II: Express Briefs}, 2023.

\bibitem{SIS2022_ISCAS}
A.~J. Leigh, M.~Heidarpur, and M.~Mirhassani, ``Selective input sparsity in
  spiking neural networks for pattern classification,'' in {\em 2022 IEEE
  International Symposium on Circuits and Systems (ISCAS)}, pp.~799--803, IEEE,
  2022.

\bibitem{cai2003transaction}
L.~Cai and D.~Gajski, ``Transaction level modeling: an overview,'' in {\em
  First IEEE/ACM/IFIP International Conference on Hardware/Software Codesign
  and Systems Synthesis (IEEE Cat. No. 03TH8721)}, pp.~19--24, IEEE, 2003.

\bibitem{kandel2000principles}
E.~R. Kandel, J.~H. Schwartz, T.~M. Jessell, S.~Siegelbaum, A.~J. Hudspeth,
  S.~Mack, {\em et~al.}, {\em Principles of neural science}, vol.~4.
\newblock McGraw-hill New York, 2000.

\bibitem{maass1997networks}
W.~Maass, ``Networks of spiking neurons: the third generation of neural network
  models,'' {\em Neural networks}, vol.~10, no.~9, pp.~1659--1671, 1997.

\bibitem{adrian1926impulses}
E.~D. Adrian and Y.~Zotterman, ``The impulses produced by sensory nerve
  endings: Part 3. impulses set up by touch and pressure,'' {\em The Journal of
  physiology}, vol.~61, no.~4, p.~465, 1926.

\bibitem{johansson2004first}
R.~S. Johansson and I.~Birznieks, ``First spikes in ensembles of human tactile
  afferents code complex spatial fingertip events,'' {\em Nature neuroscience},
  vol.~7, no.~2, pp.~170--177, 2004.

\bibitem{izhikevich2003bursts}
E.~M. Izhikevich, N.~S. Desai, E.~C. Walcott, and F.~C. Hoppensteadt, ``Bursts
  as a unit of neural information: selective communication via resonance,''
  {\em Trends in neurosciences}, vol.~26, no.~3, pp.~161--167, 2003.

\bibitem{kim2018deep}
J.~Kim, H.~Kim, S.~Huh, J.~Lee, and K.~Choi, ``Deep neural networks with
  weighted spikes,'' {\em Neurocomputing}, vol.~311, pp.~373--386, 2018.

\bibitem{arbib2003handbook}
M.~A. Arbib {\em et~al.}, ``The handbook of brain theory and neural networks,''
  tech. rep., MIT Press, 2003.

\bibitem{rathi2021diet}
N.~Rathi and K.~Roy, ``Diet-snn: A low-latency spiking neural network with
  direct input encoding and leakage and threshold optimization,'' {\em IEEE
  Transactions on Neural Networks and Learning Systems}, 2021.

\bibitem{rueckauer2017conversion}
B.~Rueckauer, I.-A. Lungu, Y.~Hu, M.~Pfeiffer, and S.-C. Liu, ``Conversion of
  continuous-valued deep networks to efficient event-driven networks for image
  classification,'' {\em Frontiers in neuroscience}, vol.~11, p.~682, 2017.

\bibitem{pasricha2002transaction}
S.~Pasricha {\em et~al.}, ``Transaction level modeling of soc with systemc
  2.0,'' in {\em Synopsys user group conference (SNUG)}, vol.~3, p.~3, 2002.

\bibitem{clouard2002towards}
A.~Clouard, G.~Mastrorocco, F.~Carbognani, A.~Perrin, and F.~Ghenassia,
  ``Towards bridging the precision gap between soc transactional and
  cycle-accurate levels,'' in {\em Proc. Design, Automation and Test in Europe
  (DATE)}, 2002.

\bibitem{mnist}
Y.~LeCun, L.~Bottou, Y.~Bengio, and P.~Haffner, ``Gradient-based learning
  applied to document recognition,'' {\em Proceedings of the IEEE}, vol.~86,
  no.~11, pp.~2278--2324, 1998.

\bibitem{fmnist}
H.~Xiao, K.~Rasul, and R.~Vollgraf, ``Fashion-mnist: a novel image dataset for
  benchmarking machine learning algorithms,'' {\em arXiv preprint
  arXiv:1708.07747}, 2017.

\bibitem{panda2001systemc}
P.~R. Panda, ``Systemc: a modeling platform supporting multiple design
  abstractions,'' in {\em Proceedings of the 14th international symposium on
  Systems synthesis}, pp.~75--80, 2001.

\bibitem{spike_conv_propose_2018}
R.~Tapiador-Morales, A.~Linares-Barranco, A.~Jimenez-Fernandez, and
  G.~Jimenez-Moreno, ``Neuromorphic lif row-by-row multiconvolution processor
  for fpga,'' {\em IEEE transactions on biomedical circuits and systems},
  vol.~13, no.~1, pp.~159--169, 2018.

\bibitem{event_conv_2022}
J.~Sommer, M.~A. {\"O}zkan, O.~Keszocze, and J.~Teich, ``Efficient hardware
  acceleration of sparsely active convolutional spiking neural networks,'' {\em
  IEEE Transactions on Computer-Aided Design of Integrated Circuits and
  Systems}, vol.~41, no.~11, pp.~3767--3778, 2022.

\bibitem{FPGANHAP}
Y.~Liu, Y.~Chen, W.~Ye, and Y.~Gui, ``Fpga-nhap: A general fpga-based
  neuromorphic hardware acceleration platform with high speed and low power,''
  {\em IEEE Transactions on Circuits and Systems I: Regular Papers}, vol.~69,
  no.~6, pp.~2553--2566, 2022.

\bibitem{ye2022implementation}
W.~Ye, Y.~Chen, and Y.~Liu, ``The implementation and optimization of
  neuromorphic hardware for supporting spiking neural networks with mlp and cnn
  topologies,'' {\em IEEE Transactions on Computer-Aided Design of Integrated
  Circuits and Systems}, 2022.

\bibitem{di2022sne_dvsgest}
A.~Di~Mauro, A.~S. Prasad, Z.~Huang, M.~Spallanzani, F.~Conti, and L.~Benini,
  ``Sne: an energy-proportional digital accelerator for sparse event-based
  convolutions,'' in {\em 2022 Design, Automation \& Test in Europe Conference
  \& Exhibition (DATE)}, pp.~825--830, IEEE, 2022.

\bibitem{li2021differentiable}
Y.~Li, Y.~Guo, S.~Zhang, S.~Deng, Y.~Hai, and S.~Gu, ``Differentiable spike:
  Rethinking gradient-descent for training spiking neural networks,'' {\em
  Advances in Neural Information Processing Systems}, vol.~34,
  pp.~23426--23439, 2021.

\bibitem{temp_vs_spatial_coding}
Z.~Pan, J.~Wu, M.~Zhang, H.~Li, and Y.~Chua, ``Neural population coding for
  effective temporal classification,'' in {\em 2019 International Joint
  Conference on Neural Networks (IJCNN)}, pp.~1--8, IEEE, 2019.

\bibitem{pop_coding_brain}
T.~D. Sanger, ``Neural population codes,'' {\em Current opinion in
  neurobiology}, vol.~13, no.~2, pp.~238--249, 2003.

\bibitem{lee2016training}
J.~H. Lee, T.~Delbruck, and M.~Pfeiffer, ``Training deep spiking neural
  networks using backpropagation,'' {\em Frontiers in neuroscience}, vol.~10,
  p.~508, 2016.

\end{thebibliography}
}
\vspace{-35pt}
\begin{IEEEbiography}
[{\includegraphics[width=1in,height=1.25in,clip,keepaspectratio]{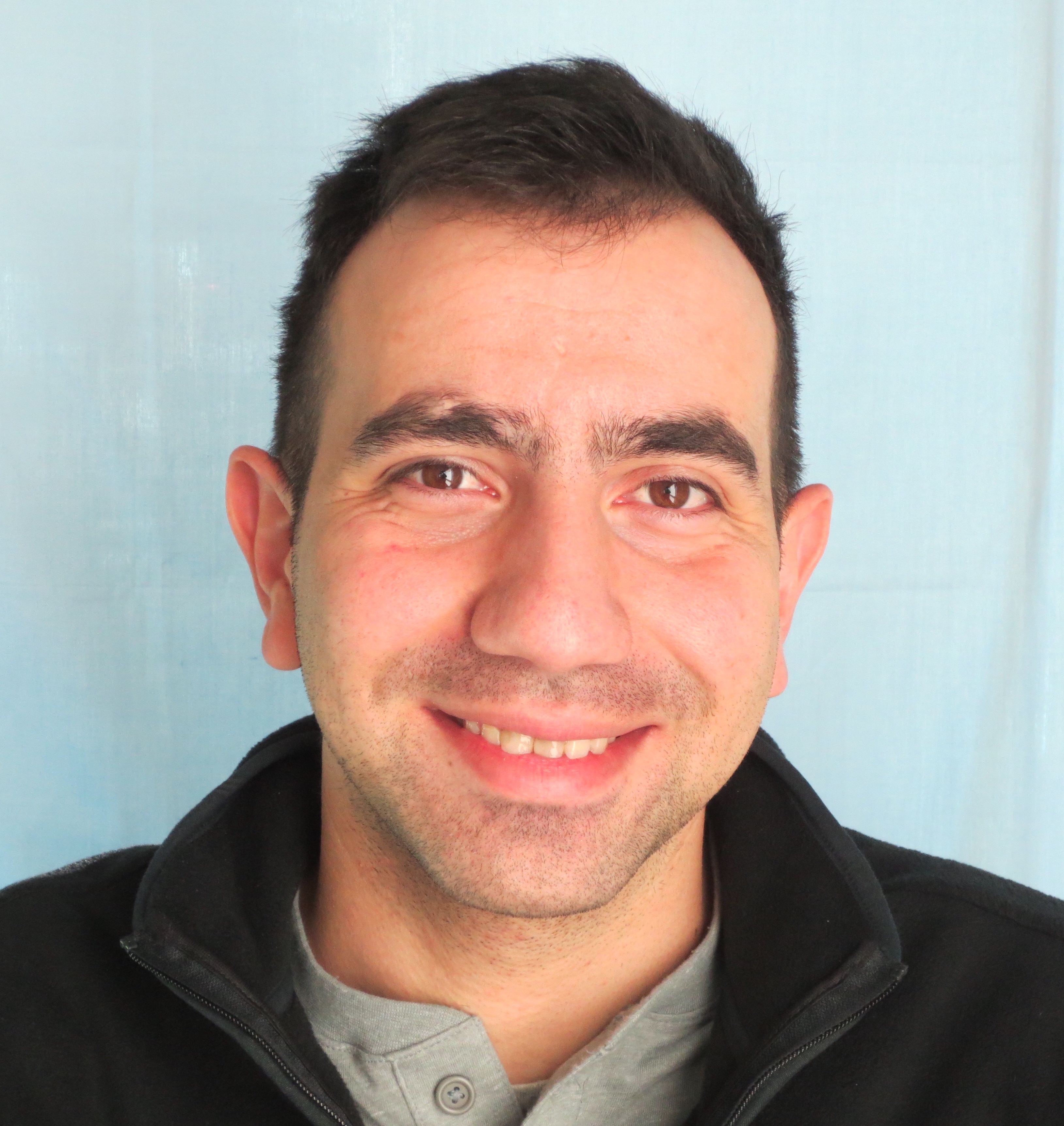}}]{Ilkin Aliyev} is a Ph.D. candidate in the Department of Electrical and Computer Engineering at the University of Arizona. He received his M.S in Electrical and Electronics Engineering from Ozyegin University in 2021 and B.S. in Electronics and Communications Engineer from Yildiz Technical University, Turkey in 2018. His research interests include energy-efficient ML accelerator design and brain-inspired computing architectures. 
\end{IEEEbiography}
\vspace{-35pt}
\begin{IEEEbiography}
[{\includegraphics[width=1in,height=1.25in,clip,keepaspectratio]{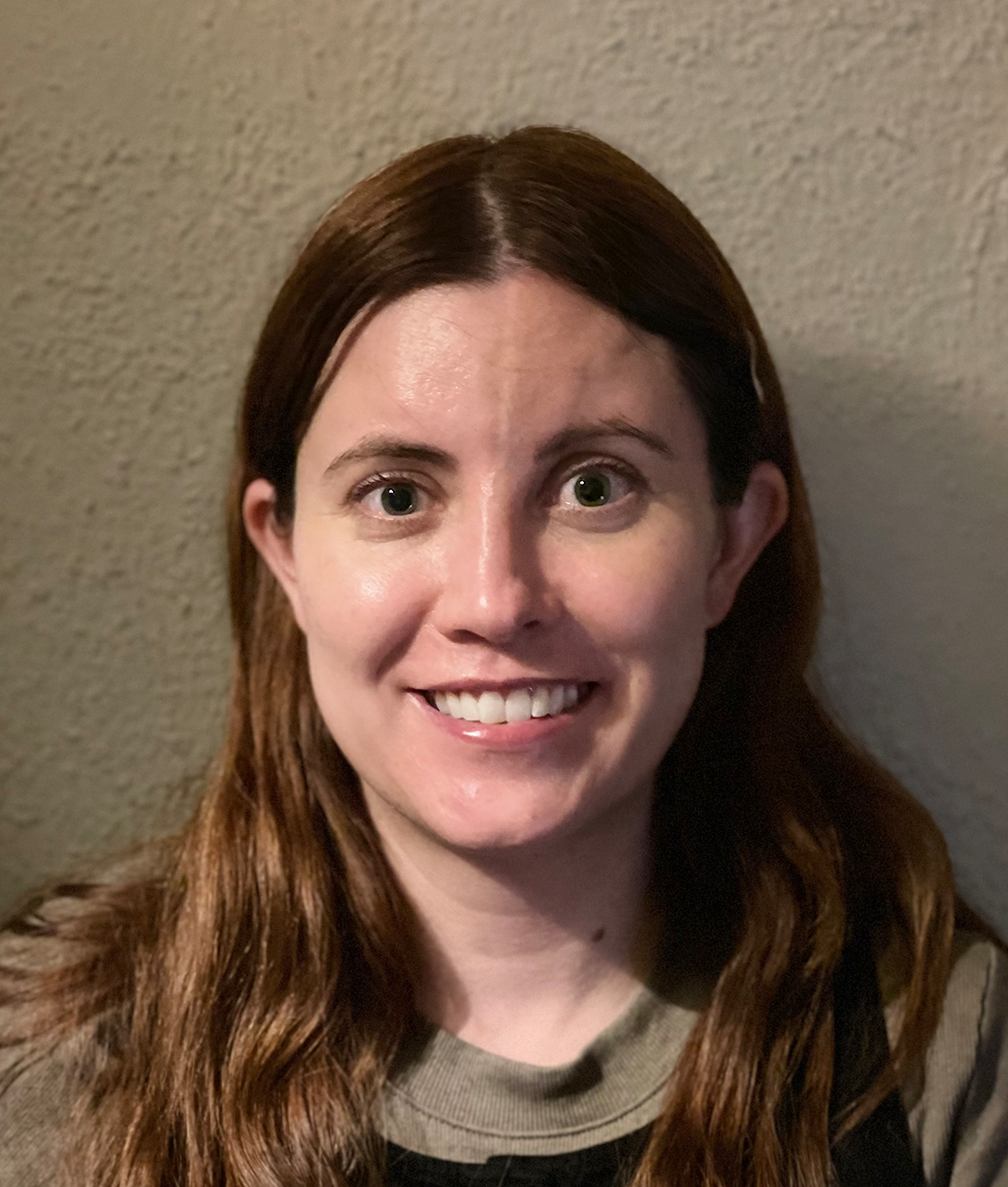}}]{Kama Svoboda} is a Ph.D. student in the Department of Electrical and Computer Engineering at the University of Arizona. She received her B.S. and M.S in Electrical and Computer Engineering from the University of Arizona in 2022 and 2023, respectively. Her research interests include computer architecture and neuromorphic computing, specifically hardware implementations of spiking neural networks.
\end{IEEEbiography}
\vspace{-35pt}
\begin{IEEEbiography}
[{\includegraphics[width=1in,height=1.25in,clip,keepaspectratio]{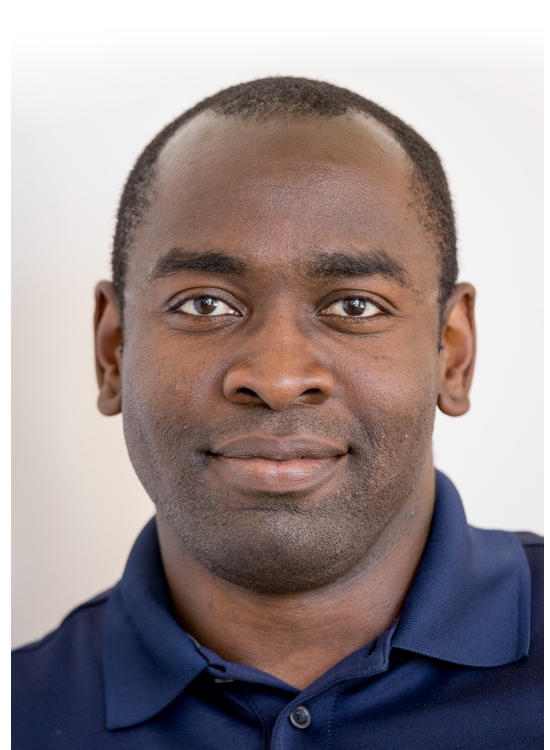}}]{Tosiron Adegbija} received his M.S and Ph.D. in Electrical and Computer Engineering from the University of Florida in 2011 and 2015, respectively and his B.Eng in Electrical Engineering from the University of Ilorin, Nigeria in 2005.
He is currently an Associate Professor of Electrical and Computer Engineering at the University of Arizona, USA. His research interests are in computer architecture, with an emphasis on brain-inspired computing, adaptable computing, low-power embedded systems design and optimization methodologies, and domain-specific architectures. He received the CAREER Award from the National Science Foundation in 2019. He is a Senior Member of the IEEE.
\end{IEEEbiography}

\end{document}